\documentclass[preprint2]{aastex}

\slugcomment{Not to appear in Nonlearned J., 45.}

\shorttitle{A Galaxy Model from 2MASS Star Counts}
\shortauthors{Polido et al.}

\begin{document}


\title{A Galaxy Model from 2MASS Star Counts in the Whole Sky Including the Plane}

\author{P. Polido\altaffilmark{1} and F. Jablonski\altaffilmark{2}}
\affil{Divis\~ao de Astrof\'{\i}sica, Instituto Nacional de Pesquisas Espaciais, Avenida dos Astronautas 1758, 12227-010 S\~ao Jos\'e dos Campos SP, Brazil}

\and

\author{J. R. D. L\'epine\altaffilmark{3}}
\affil{Instituto de Astronomia, Geof\'{\i}sica e Ci\^encias Atmosf\'ericas, Universidade de S\~ao Paulo, Rua do Mat\~ao 1226, 05508-900 S\~ao Paulo SP, Brazil}


\altaffiltext{1}{pripolido@gmail.com}


\begin{abstract}

We use the star counts model of Ortiz \& L\'epine (1993) to perform an unprecedented exploration of the most important 
galactic parameters comparing the predicted counts with the 2MASS observed star counts in the J, H and K$_{S}$ bands 
for a grid of positions covering the whole sky. The comparison is made using a grid of lines-of-sight given by the 
HEALPix pixelization scheme. The resulting best fit values for the parameters are: (2120$\pm$200) pc for the radial scale length and (205$\pm$40) pc for the scale height of the thin disk, with a central hole of (2070$_{-800}^{+2000}$) pc for the same disk; (3050$\pm$500) pc for the radial scale length and  (640$\pm$70) pc for the scale height of the thick disk; (400$\pm$100) pc for the central dimension of the spheroid, (0.0082$\pm$0.0030) for the spheroid 
to disk density ratio, and (0.57$\pm$0.05) for the oblate spheroid parameter.

\end{abstract}



\section{Introduction}

The star counts method is an important tool for the investigation of the galactic structure and is based on the equation 
of stellar statistics \citep{binney}:

\begin{eqnarray}
N_{s}(m_{1},m_{2},l,b)d\Omega = \int^{m_{2}}_{m_{1}} dm \nonumber \\ \int_{0}^{\infty} r^{2}dr \rho_{s}(r,M) \phi_{s}(M) d\Omega,
\label{eq:estatis}
\end{eqnarray}

\noindent which allows us to predict the number of objects for a certain line-of-sight defined by the galactic 
longitude $l$ and the galactic latitude $b$. $N_{s}$ is the number of stars of type $s$ with apparent magnitude 
between $m_{1}$ and $m_{2}$ in a solid angle $d\Omega$ in the direction defined by $(l,b)$; $r$ is 
the heliocentric distance, $\rho_{s}$ is the stellar density and $\phi_{s}$ is the luminosity function. 
We adopted the luminosity function as the number of stars per cubic parsec near the Sun with magnitudes in 
the range $(M, M + dM)$. It is convenient to handle the stellar densities normalized with respect to the values in the solar neighborhood.

The star counts models can be classified in roughly two categories: those adopting an empirical luminosity 
function -- based on observations of the stellar populations in the solar neighborhood and specific environments, like
globular clusters -- and those adopting a 
luminosity function derived from theoretical stellar evolutionary tracks and the associated distributions 
of stellar masses,  ages and metallicities. The pioneer investigation of \citet{bahcallsoneira} fits into 
the first group, together with models like the ones by \citet{wainscoat}, \citet{ortizlepine} 
(hereafter OL93), \citet{juric2008} and \citet{chang}. On the other hand, the works of \citet{besancon} 
and \citet{girardi2005} belong to the group of models which adopt a stellar evolution approach to 
define the luminosity function.

The number of components used to represent the Galaxy varies from model to model, according to the focus of the 
work. Basically, two components are always present: a disk and a spheroid. After \citet{gilmore1983} 
many authors adopted the disk component consisting of a thin disk and a thick disk. 
Concerning the spheroid, despite this terminology  being used in all models, it actually may indicate 
different galactic structures, like the bulge, the halo or even both (e.g., OL93). The densities 
of stars on the disks decrease exponentially with galactocentric radius and with the vertical distance 
to the plane, whereas the density in the spheroid follows a decay similar to the de Vaucouleurs' law.

Despite strong evidence that our galaxy has a spiral structure, the number of arms, 
as well as the parameters which describe their shapes and positions vary from work to work (e.g., \citet{russeil2003,
levine2006, vallee2008, hou2009}). The star counts models which attempt to take the spiral arms into account are 
those due to OL93 and \citet{wainscoat}.

The existence of a bar in the central region of our galaxy was proposed in the 1970s based on the large non-circular 
motions seen in the observations of HI and CO in the inner Galaxy \citep{peters1975,cohen1976,liszt1980}, 
but only in the 1990s the combination of evidences such as the NIR light distribution \citep{blitz1991,weiland1994}, 
star counts asymmetries \citep{nakada1991,stanek1997,benjamin2005}, gas kinematics \citep{binney1991,englmaier1999,fux1999} 
and large microlensing optical depth \citep{udalski1994,zhao1995,han1995} became more persuasive.

\citet{bahcallsoneira} examined the star counts in the BV bands and assumed that the Galaxy could be well 
represented by two components: a disk and a spheroid. \citet{besancon} examined the star counts in the UBV bands 
and used three components to model the Galaxy: a disk, a halo and an exponential spheroid population with intermediate age. 
\citet{wainscoat} adopted five components (disk, halo, bulge, spiral arms, molecular ring) to model 
the Galaxy star counts as a foreground for their extragalactic source counts based on Infrared Astronomical 
Satellite Survey (IRAS)\footnote{http://irsa.ipac.caltech.edu/Missions/iras.html}. OL93 attempted to model 
the star counts in RIJHKL, [12 $\mu$m] and [25 $\mu$m] IRAS bands, using four components: a spheroid, 
two disks with scale heights 100 pc and 390 pc, and spiral arms. A version of the model 
with an intermediate disk and a bar is described in their 
website\footnote {http://www.astro.iag.usp.br/$\sim$jacques/pingas.html}. 
The approach of \citet{girardi2005} uses a stellar populations synthesis code as the input for the luminosity 
function to compare their four components model (thick disk, thin disk, halo, bulge) with optical and infrared 
data, including the Two Micron All Sky Survey (2MASS)\footnote{http://www.ipac.caltech.edu/2mass/}, 
Chandra Deep Field South (CDFS) and Hipparcos\footnote{http://www.rssd.esa.int/Hipparcos/}. \citet{juric2008} 
have compared the stars counts from the Sloan Digital Sky Survey (SDSS)\footnote{http://www.sdss.org/} with 
the predicted counts from their galactic model with a disk and a halo. \citet{chang} used a model with a 
spheroid and a disk, and a power-law luminosity function for which the power index is a free parameter at each grid 
point on the sky for galactic latitudes $|b| > 30^\circ$. The comparison was made for 8192 lines-of-sight 
in the 2MASS K$_S$ band.

In this study we use the homogeneous all sky coverage of the 2MASS to test the OL93 model in the JHK$_S$ bands. 
We explore the parameters space of this model adopting an updated luminosity function to describe the stellar populations in the solar neighborhood.

In sections 2 and 3, we describe the OL93 model and the method used to search for the best fit parameters. The results 
and discussion are presented in section 4. Our conclusions follow in the last section.

\section{The Galactic Model}
\label{section:modelo}

The OL93 model uses a  young disk, an old disk, an intermediate disk composed of C-stars, a spheroid, spiral arms and a bar,  
with the total number density of sources being the sum of all individual contributions. A detailed description of  
all components follows.

\subsection{Disks}

We consider a thin (young) disk and a thick (old) disk. The density of each subcomponent $i=Y{(\rm oung)}, O{\rm (ld)}$, 
is given by a modified exponential \citep{leroy2000}, which is an important change with respect to the original model:

\begin{equation}
n_{d,i}(r, z, s) = n_{d,i} (R_{0}, 0, s) e^ {- \frac{r}{\alpha_i}- \frac{\beta_i}{r}- \frac{z}{h_{i}(r)}}.
\end{equation}
 
\noindent Here $r$ is the distance to the galactic center in the plane of the Galaxy,  $n_{d,i} (R_{0}, 0, s)$ is the star 
density of spectral type $s$ in the solar neighborhood, $\alpha_i$ is the radial scale length, which does not depend 
on the spectral type, 
$h_{i}(r)$ is the scale height and $\beta_{i}$ is the radius of a hole in the central part of the disk. The thick disk 
has a distribution of spectral types of giants similar to that found in globular clusters. All the stars of spectral 
types O and B, as well as the supergiants of all spectral types were excluded from de composition of the thick disk. 
The scale height $h_{i}(r)$ is given by:

\begin{equation}
h_{i}(r) = z_{i}  e^{\frac{0.4 \left( r - R_{0} \right) }{R_{0}}},
\label{zxr}
\end{equation}

\noindent where $z_{i}$ is the scale height in the solar neighborhood.  The free parameters of the model related to
the disks are $\alpha_Y, \alpha_O, \beta_Y, \beta_O, z_Y {~\rm and~} z_O$.

Recent studies (e.g., \citet{sofue2009} and \citet{amores2009}) suggest the existence of a 
minimum in the density of young objects in the solar neighborhood due to the effects of corotation of the spiral
pattern with the galactic rotation. We experimented with this possibility modeling the young disk density with

\begin{equation}
n_{d,Y}(r) = n_{d,Y} (R_0) \left[  1.3-0.3 e^{-\left(r-R_0\right)^{2}}\right],
\label{densjacques}
\end{equation}

\noindent where $r$ and $R_0$ are expressed in kpc. The results of this attempt are shown in section 4.2.1.

OL93 also consider an intermediate disk consisting of carbon stars which are bright in infrared mainly during their 
Asymptotic Giant Branch (AGB) phase, when they lose mass at high rates. This component does not show radial dependence 
\citep{guglielmo1998} but its density decreases exponentially with the distance to the galactic plane with a scale height 
of 200 pc. We did not attempt determining any parameter related to this component.

The OL93 model does not use warp components on the disks. Evidences for the presence of this contribution in 2MASS data are presented by \citet{lopezcorredoira2002} and \citet{reyle2009}. Uncertainties related to the different space distributions of dust and stars \citep{freuden1994,drimmel2001,reyle2009} and differences in the distribution of giant and main sequence stars (eg. \citet{lopezcorredoira2002}) led us not to attempt including this component in our modeling.

\subsection{Spheroid}

The spheroidal structure is subtle, and when we observe it, specially close to the galactic center, we see the superimposition of contributions from the young and old disks (which may or may not have central holes), and  a galactic bar. Besides, the innermost regions of the spheroid are subject to heavy reddening. A number of improvements to the description of this region as a whole have been suggested in the last decade. \citet{lopezcorredoira2005} found that the best fit to 2MASS data is given by a structure  which consists of a boxy bar associated to a tri-axial bulge. \citet{vanholle2009} also use a tri-axial figure which is truncated by a Gaussian decay, and \citet{robin2012} describe the 2MASS data in the central region of the galaxy ($|l| < 20^\circ, |b| < 10^\circ$) with a triaxial boxy bar/bulge plus a longer and thicker ellipsoid. \citet{nataf2012}, \citet{mcwilliam2010} and \citet{saito2011} investigated  the  X-shaped component which is part of the bar/bulge structure.

The OL93 model uses an oblate spheroid in an attempt to describe both the inner region and larger outer scales. The density in the spheroid is adapted from the mass density of \citet{hernquist1990} and is given by:
\begin{equation}
n_{sph}(R,s) = \frac{C_1}{\zeta (\zeta + a_H)^{3}} ,
\label{spheroid}
\end{equation}

\noindent where $R$ is the distance to the galactic center, $\zeta = \sqrt{(z/\kappa)^2 + R^2}$, with $\kappa$ being
the oblateness of the spheroid and $z$ the distance above the galactic plane.  $a_H$ is a scale length, and 
$C_1 = R_0(R_0 + a_H)^3 (N_{sph}/N_D)$. The last term is the ratio of densities of spheroid and disk populations in 
the solar neighborhood. $R_0$ is the distance of the sun to the galactic center, taken here as 8.0\,kpc -- very close to both arithmetic and weighted means of $R_0$ determinations obtained since 1992 (\citet{malkin2013,morris2012,reid2012,reid2013,zhu2013}). OL93 used R$_{0}=7.9$\,kpc, which was intermediate between the IAU 
recommended 8.5 kpc and the ``short scale"  value 7.5 kpc often used at the time (see for instance a discussion in Section 2 of \citet{lepine2011b}).
The stellar population of the spheroid is considered the same as that of the thick disk, discussed below.
$a_H$, $\kappa$ and $N_{sph}/N_D$ are free parameters in the OL93 model.

\subsection{Spiral arms}

One of the first representations of the spiral structure in the Milky Way, due to \citet{georgelin}, proposed a four arms pattern. Since then, many authors have studied this galactic component, obtaining discrepant results for the number of arms and their locations. Recent
results include  \citet{majaess2009}, based on the space distribution of type II Cepheids which suggest deviations from what logarithmic spirals 
predict for the Saggitarius-Carina arm and the Local arm. \citet{lepine2011b} traced the CS molecular emission related to IRAS sources and
concluded that they are not fitted by logarithmical arms, rather by a sequence of straight line segments. From 2MASS data and gas distributions, \citet{francis2012} concluded that the Milky Way is a two-armed grand-design  spiral. \citet{robitaille2012} found the existence of two dominant and two secondary arms on GLIMPSE, MIPSGAL and IRAS data. In the outer parts of the Galaxy the spiral arms are not very prominent. \citet{quillen2002} even suggested, based on 2MASS data, that the Galaxy is flocculent in that region. Comprehensive reviews of earlier works can be found in \citet{vallee1995,vallee2002,vallee2005}.

The spiral pattern is the same as in the OL93 model, with four logarithmic arms, each described by:

\begin{equation}
r_{arm} = q e^{(\theta - \theta_{0})tan(i)} ,
\label{arms}
\end{equation}

\noindent where $q$ is the galactic radius where each arm begins, $\theta_{0}$ is the initial galactocentric angle and $i$ is 
the pitch angle. 
The arms are 
confined to the range of galactic radii 2 kpc $< r <$ 15 kpc and consist of O5-B stars of all luminosity 
classes and supergiants of all spectral types. We adopted an improved version of the OL93 arms configuration, 
consistent with recent observations. Values for $q$, $i$ and  
$\theta_{0}$ were derived from the fit of Equation \ref{arms} to the representation seen in \citet{churchwell2009} which is based on the model of \citet{georgelin} with some modifications: variations in positions according 
to results of parallax distances to masers in regions of stellar formation \citep{xu2006}; refinement in the 
tangential directions to the arms from CO surveys \citep{dame2001}; revisions in the amplitudes of the arms due 
to GLIMPSE results and the work of \citet{drimmel2001}; and finally, the locations of the outer and distant arms using kinematical 
data \citep{mcclure2004}.
 
The only parameter related to the spiral pattern which we attempted to derive from the 2MASS observations is the density contrast between the arms and the thin disk, $C_S$, 
since the arms behave as enhancements of the disk density. The stars density perpendicular to the 
arm's length is described by a Gaussian function and its half width at half maximum is taken as $\sim 180$\,pc.
The tangential directions to the arms, which are also the directions where we expect to see the 
largest contribution of the arms to the star counts, are located at longitudes 32$^{\circ}$, 49$^{\circ}$, 284$^{\circ}$ and 308$^{\circ}$.

\subsection{Bar}

Prior information on the structural properties of the bar is somewhat sparse. It is generally accepted that our galaxy has a bar, but its length, shape and orientation cover wide ranges in the literature.  The half length of the bar, for example, ranges from 0.67 kpc in \citet{cao2013} to 3.9 kpc in \citet{lopezcorredoira2007}.

In our modeling, the bar contribution to the stellar density is given by:

\begin{eqnarray}
   r_{bar}    & = & \left\lbrace  \left[ \left(\frac{x^\prime}{x_0}\right)^2 + \left(\frac{y^\prime}{y_0}\right)^2 \right]^2 + \left( \frac{z^\prime}{z_0} \right)^4 \right\rbrace ^{\frac{1}{4}} \nonumber \\
   n_{bar} & = & (n_{d,Y} + n_{d,O}) C_{bar}\, e^{-\frac{1}{2} r_{bar}^2}
\end{eqnarray}

\noindent where $r_{bar}$ describes the shape ($x^\prime, y^\prime, z^\prime$ are along the three axes) and $C_{bar}$ gives the contribution of the bar relative to the sum of the disk densities \citep{dwek1995}. We explored axes ratios in the ranges $\left\lbrace x_{0}:y_{0}:z_{0}\right\rbrace =\left\lbrace 1.00:(0.22-0.67):(0.34-0.40)\right\rbrace $ and $x_0$ up to 4 kpc, a little more than the largest value found in the literature \citep{lopezcorredoira2007}. The orientation of the bar with respect to the line Sun-GC, $\theta_{bar}$, was allowed to be in the range $11^\circ$ -- $53^\circ$. Even though not strongly constrained by the 2MASS data, we attempted to determine the geometrical parameters of the bar as well as the density contrast between the bar and the disks.

\subsection{Interstellar extinction}

The interstellar extinction is an important ingredient of a galactic model, specially in the 
galactic plane. Due to the accumulation of dust in the galactic plane, the extinction is larger
in this region of the Galaxy. The model of \citet{amores} is based on the distribution of gas (HI and CO) and 
interstellar dust (IRAS 100 $\mu$m), under the assumption that the dust is well mixed with the gas. In this model, 
the Galaxy has axial symmetry, the gas density varies radially in a smooth way and the spiral arms produce no 
effects in the extinction. The interstellar extinction is calculated assuming that it is proportional to 
column density of hydrogen, in both atomic ($N_{HI}$) and molecular ($N_{H_{2}}$) forms:

\begin{equation}
A_{V} = C_V(r) N_{HI}(R,z) + 2 C_V(r) N_{H_{2}}(R,z) .
\end{equation}

\noindent Here $C_V$ is a proportionality factor, with average value of $5.3 \times 10^{-22}$ mag cm$^{2}$ \citep{bohlin}, 
if $A_{V}=3.1E_{B-V}$, but it can change along the galactocentric radius due to the metallicity gradient. 
For galactocentric 
distances in the plane with $r > 1.2$ kpc, the authors suggest a proportionality with $r^{-0.5}$. For $r < 1.2$ kpc, due 
to the highly uncertain metallicity of the region, a constant value is employed. Similar analytical expressions were 
adopted for both gas forms:

\begin{equation}
n_{HI,H_{2}}=c e^{- \frac{r}{a} -\left( \frac{b}{r}\right)^{2}} ,
\end{equation}

\noindent where, for HI, $a=$7 kpc, $b=$1.9 kpc and $c=$0.7 $cm^{-3}$, and for H$_{2}$, $a=$1.2 kpc, $b=$3.5 kpc 
and $c=$58 $cm^{-3}$. Since there is a large concentration of H$_{2}$ in the galactic center, for $r <$ 1.2 kpc this 
was modeled separately with the function

\begin{equation}
n_{H_{2}}= d e^{-\left(\frac{r}{f}\right)^{2}} ,
\end{equation}

\noindent with $f = 0.1$ kpc and $d = 240 $cm$^{-3}$.

The vertical distribution of hydrogen is given by a Gaussian function of $z$:

\begin{equation}
n_{H} (r,z) =  n_{H} (r) e^{- \frac{1}{2} \frac{z^2}{(1.18 z_{1/2})^{2}}} ,
\end{equation}

\noindent where $z_{1/2}$ is the half width at half height of the scale height.
\citet{amores} suggest for the scale height of H$_{2}$:

\begin{equation}
z_{1/2} = 45 e^{0.1r} \rm pc,
\end{equation}

\noindent while for HI the same expression must be multiplied by a factor 1.8.

\subsection{Luminosity function}

\begin{figure}
 \includegraphics[scale=0.7,angle=270]{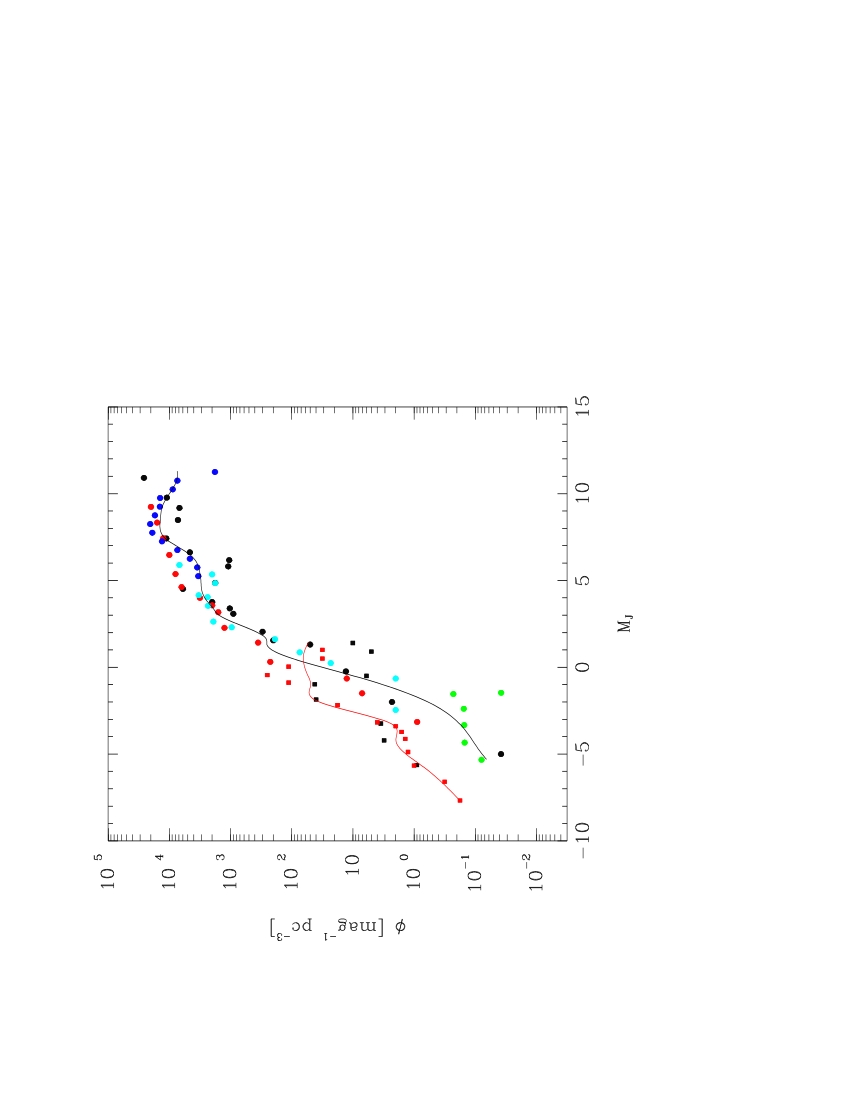}
\caption{Updated luminosity functions for main sequence and giant stars in the J band. The circles and squares indicate, respectively, the main sequence and giants luminosity functions according to \citet{ortizlepine} (black), \citet{wainscoat} (red), \citet{bochanski2010} (blue), \citet{reid1997} (green) and \citet{murray1997} (cyan). The lines indicate the smooth fits for the main sequence (black) and giants (red).\label{fig1}}. 
 \end{figure}

The luminosity function used in this work follows OL93 and enters the code via a table containing up to 64 classes of objects among 
main sequence, giants, supergiants and variable objects. The space densities were updated to be consistent with recent 
results. Figure \ref{fig1} shows 
the luminosity function for main sequence and giant stars according to OL93, \citet{wainscoat}, \citet{bochanski2010}, 
\citet{reid1997} and \citet{murray1997}. The luminosity functions for supergiants and variable objects are the same
as in OL93.

\section{Methodology} \label{section:metod}
\subsection{The data: star counts from the 2MASS catalog} \label{dados2mass}

Since  interstellar extinction is lower in the near infrared (NIR), surveys in this region are efficient  to investigate the structure of the Galaxy. The 2MASS project \citep{skrut,cutri2003} covers 99.8\% of the sky in the 
NIR with 471 million point sources observed in  
J (1.25 $\mu$m), H (1.65 $\mu$m) and K$_{S}$ (2.17 $\mu$m).  At SNR=10, the limit magnitudes for  
point objects are 15.8, 15.1 and 14.3 in the J, H and K$_{S}$ bands, respectively. These limits refer to
unconfused sources outside the galactic plane ($|b| > 10^{\circ}$) and far from areas where the interstellar extinction is large.
The all-sky coverage of the 2MASS data allows us to make a comprehensive comparison of the observed counts with the predictions of the model by OL93.

The number of sources in a given line-of-sight through the Galaxy can be obtained from the 2MASS database in 
three forms: cone, box and polygon search. Taking into account practical aspects such as the elapsed time 
for retrieving the data, we opted to use the cone search, with a cone area of one square degree. We use a grid 
of galactic longitudes and latitudes as generated by the Hierarchical Equal Area Isolatitude pixelization of 
the Sphere (HEALPix)\footnote{ http://healpix.jpl.nasa.gov/} \citep{gorski} 
in order to obtain a uniform sampling in galactic coordinates. The orientation of the grid is such that the 
``equator" of the HEALPix scheme coincides with the galactic plane. Our basic grid contains 192 points (the N$_{side}$ parameter of HEALPix = 4). A N$_{side}$=8 grid (768 points) was used to provide four neighbors to each point of the N$_{side}$=4 scheme in order to estimate the variance of the star counts at each position of the basic grid. This allows us to verify if the assumption of using a  $\sqrt{N}$ law for the uncertainty in the counts, $N$, is reasonable. 
It turns out that the dispersion obtained from the five points estimate is consistent with the Poisson 
uncertainties for most of the sky, but significant differences show up close to the galactic plane.  
This simply reflects the structure of the Galaxy in a scale of a few degrees and the large gradients in star counts along the
$z$ direction close to the galactic plane. We investigated the results of using one or other option for the uncertainties in
the observed star counts and the results are shown in section \ref{section:results}.

The output of a cone search is a list of JHK$_S$ magnitudes and catalog flags for each counted object. The flags include important indicators to eliminate low quality counts. We examined the following flags: photometric quality, read flag, 
contamination and confusion flag, and galactic contamination (extended objects) flag. Table \ref{tb:tb1}
summarizes the criteria to consider a source \textit{bad}. Table \ref{tb:tb2} summarizes the percentage of rejections for a few 
lines-of-sight. One can see that even far from the galactic plane, the K$_S$ band is more prone to be affected by rejections 
than the other bands, while in regions near the galactic plane the three bands have about 50\% rejections. 
This indicates that comparisons in the galactic plane should not go to magnitudes fainter than 11. Using an extreme 
field as an example ($l = 11.25^{\circ}, b = 0.0^{\circ}$), at the 11th magnitude bin we find 74 rejections out of a total of 
1062 counts for the J band, 566 (in a total of 4860) for the H band and 2076 (in a total of 9990) for the K$_{S}$ band (7\%, 
11.6\% and 20.8\%, respectively). The fractions of rejections are 25 to 30\% from photometric quality, 10-15\% from 
the read flag and 65-75\% from the contamination/confusion flag.

The limiting magnitude varies from grid point to grid point due to differences in interstellar extinction and to
the existence of non-resolved close sources. This is specially true for regions in the galactic plane and close to the 
galactic center. To set up a limiting magnitude for each grid point we stepped in magnitude, counting the corresponding 
objects and monitoring the number of associated rejections. The limiting magnitude was set when the fraction of rejected 
objects reached 10\%.

Objects not rejected are put into ascending order of magnitude in each band and the 
cumulative star counts are obtained from the sums up to a certain limit magnitude. The differential 
counts can be obtained from the latter.

\subsection{Parameters estimation}

Models in which multiple parameters are to be evaluated are recognizably challenging, in the sense that traditional
methods of search for an optimal solution, like those based on gradients in the figure of merit (e.g., 
the Downhill Simplex method of \citet{simplex}) may lead to
local maxima (or minima) which can be far from the best solution. A number of statistical methods has received attention in the last decades because even though being relatively slow, they produce reliable results. We chose to use the Markov chain Monte Carlo method to have an overall view of the parameter space, including a rough localization but with a reliable measure of the spread of the parameters, and the Nested Sampling (NS) method to have a better estimate of the parameter values. We experimented with MCMC chains of up to $10^5$ iterations to have an overview
of possible multiple modes in parameter space. Our conclusion is that with typically $10^4$ iterations one can unambiguously limit the region of interest for each parameter. The NS procedure subsequently progresses faster to the mode, in some cases with only a few hundred iterations.

\subsubsection{MCMC}

Markov chain Monte Carlo (MCMC) \citep{gilks1996} is probably the preferred first approach for an overall view of
the parameter space in a multi-parameter problem. It has the virtue of being very simple to code and 
provides a first assessment of the location and spread of the parameters. It is perfectly suited for the Bayesian context, where prior information may be relevant in parameter estimation or model selection. Briefly, if we have a set of data $D$ with individual points $d_1, d_2,..., d_N$ and a model $M$ with a vector of parameters 
$\theta_1, \theta_2,...,\theta_{N_p}$, 
for which we are able to calculate the likelihood $\mathcal{L}$, the MCMC algorithm progresses as follows.

\begin{enumerate}
 \item Consider an initial state $\theta$, randomly chosen, with associated likelihood  $\mathcal{L}$.
 \item Draw a random candidate state $\theta^*$, statistically centered on $\theta$, for which the likelihood is $\mathcal{L}^*$.
 \item Calculate the ratio $\alpha = \frac{\mathcal{L}^*}{\mathcal{L}}$

 \item
if $\alpha > 1$ \\
   $~~~~~$ accept the new state \\
else \\
   $~~~~~$draw a random number $\beta$ in $[0,1]$\\
   $~~~~~$if $\beta < \alpha$ \\
      $~~~~~$ $~~~~~$accept the new state \\
   $~~~~~$ else \\
      $~~~~~$ $~~~~~$ stay in the old state
\item Record the chosen state and continue at step~2.
\end{enumerate}

\noindent Since we considered flat priors for all parameters, the histograms of the chain for each parameter may be regarded as proportional to the {\em a posteriori} distribution of the parameter. Each histogram location gives us an estimate of the value of the parameter itself and the confidence region (or uncertainty) may be obtained by integrating the area 
(for example, the one corresponding to the 1-$\sigma$ quantile in a normal distribution) around the mode or the median of the histogram.

By far step 2 in the scheme above is the most involving, since it implies defining a {\em step size} by which to statistically draw the proposal states $\theta^*$. Suppose a particular parameter $h$. As for all other variables,  we work on normalized quantities (here $h_n$) calculating
 $h_n = (h-h_{min})/(h_{max}-h_{min})$ and then draw the proposal $h_n^* \sim \mathcal{N}(h_n,0.289\Delta)$. 
Here ``$\sim$'' means ``distributed as''. The numerical
factor $0.289$ ensures that the step variance is similar to the variance generated by the step $\Delta$ in a uniform distribution. With this scheme, all normalized variables are subject to the same step size. We follow the recommendations in the 
literature (e.g., \citet{sivia}) and tune $\Delta$ as the chain progresses to have an acceptance rate of $\sim 37\%$.
The initial value of $\Delta$ is 0.25, and the adaptive scheme for changing its size is only started after $2^{N_p}$ steps of the chain, to allow for exploration of distant regions in parameter space.

We investigated two forms for the likelihood $\mathcal{L}$. In the first, we calculate for the whole
grid of $n_{pix}$ lines-of-sight (in one band, say H),

\begin{equation}
\chi^{2}_{\rm H}= \frac{1}{n_{pix}}\sum^{n_{pix}} \frac{1}{\rm N_{b}} \sum_{j=m_{l}}^{m_{u}} \frac{(C_{obs,j} - C_{M,j})^{2}}{(C_{obs,j} + C_{M,j})}.
\label{eq:qui2}
\end{equation}

\noindent Here $C_{obs,j}$ are the 2MASS observed counts in each line-of-sight, with $j$ running from the lower magnitude limit $m_{l}$
to the upper magnitude limit $m_{u}$ and  $C_{M,j}$ are the corresponding model quantities. Each line-of-sight has
a cumulative histogram of ${\rm N_{b}} = \left( \left[ (m_u-m_l)/\Delta m\right]  +1\right) $ magnitude bins. The other bands are treated 
similarly, and we finally write
for the likelihood

\begin{equation}
\mathcal{L} \propto \exp(-\frac{1}{2}\chi^{2})
\label{like1}
\end{equation}

\noindent with $\chi^{2} = \chi^{2}_{\rm J}+\chi^{2}_{\rm H}+\chi^{2}_{\rm K_{S}}$.

Since Equation\,\ref{like1} assumes that the $C_{obs}$ terms in Equation\,\ref{eq:qui2} are
statistically independent and normally distributed -- and the latter is not the case for low counts --
we also used the Poisson likelihood form prescribed in \citet{bienayme1987} and \citet{robin1996}:

\begin{equation}
\ln \mathcal{L} = \sum^{n_{pix}} \sum_{j=m_{l}}^{m_{u}} C_{obs,j} \left( 1 - \zeta_{j} + \ln \zeta_{j} \right),
\label{poisson}
\end{equation}

\noindent where $\zeta_{j} = \frac{C_{M,j}}{C_{obs,j}}$.

The two likelihood forms allow us to perform a sanity check on the results since Equation\,\ref{eq:qui2} gives more weight to instances of large star counts while Equation\,\ref{poisson} attributes equal weights to the counts in all magnitude bins.  The largest difference in parameter values, however, is small, typically less than  5\%. The median of the absolute value of the  relative residuals, ($(C_{obs,11}-C_{M,11})/C_{M,11}$, in a grid of 3072 lines-of-sight is typically  less than 2\%.

\subsubsection{Nested Sampling}
Nested Sampling (NS) is an algorithm for optimization in multi-parameter problems invented by \citet{skilling2004}.
It relies on the idea that whatever the number of parameters in a model, we can always populate the parameter space
with a number $N_{live}$ of random samples and calculate their likelihoods. Instead of focusing on the best likelihoods,
the algorithm works on the worst. Our implementation of NS is as follows:
\begin{enumerate}
 \item Populate the parameter space with $N_{live}$ samples chosen randomly. Notice that a minimum minimorum choice
for $N_{live}$ would be $2^{N_p}$, where $N_p$ is the number of parameters being sought. Calculate the associated 
$\mathcal{L}_i,~ i=1,...,N_{live}$. The associated $j=1,...,N_p$ parameters are stored in the vectors $\theta_{i,j}$
 \item Find the {\em worst} $\mathcal{L}$ among the $N_{live}~ \mathcal{L}_i$ and call it 
$\mathcal{L}^* = \mathcal{L}_{i_{worst}}$
 \item Choose at random an index $k$ among $1,...,N_{live}$ such that $k \ne i_{worst}$
 \item Explore the vicinity of point $\theta_k$ with a short MCMC, and choose from its
 output a parameter set $\theta_{copy}$ for which $ \mathcal{L} > \mathcal{L}^*$
 \item Substitute $\theta_{i_{worst}}$ with $\theta_{copy}$ and jump to step 2.
\end{enumerate}

A simple scheme for stopping the NS algorithm was set based on the size of the MCMC proposal step in item 4 above.
When $\Delta < 0.0001$, we finish the iterations. Again, $\Delta$ is allowed to vary only after $2^{N_p}$ steps of the NS algorithm.

The limits we adopted for parameters search are shown in Table \ref{tb:tb3}. They were chosen so to cover with some slack the range of  values found in the literature.

\subsection{Finer search for selected parameters} \label{section:finer}
\begin{figure}
 \includegraphics[scale=0.9,angle=270] {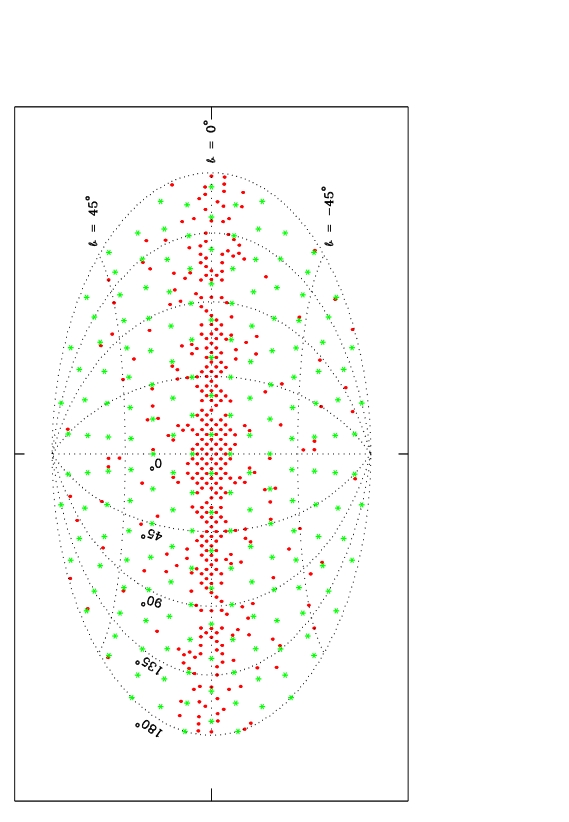}
 \caption{Illustration of the N$_{side}$=4 (192 points) basic grid in green and a finer grid with 382 points drawn from the N$_{side}$=16 HEALPix scheme (red).  \label{fig2}}. 
 \end{figure}

The N$_{side}$=4 HEALPix grid provides a good first assessment of the galactic parameters but has the obvious limitation of being too coarse, specially in regions close to the galactic center. To circumvent that, we did several experiments with finer grids, for which only a few parameters were explored. For example, a grid of 382 lines-of-sight drawn from N$_{side}$=16 HEALPix scheme provides a good coverage of the central region of the galaxy and is not very expensive in terms of computing time. At $l=b=0^\circ$ it samples every $5.6^\circ$ in longitude and $2.4^\circ$ in latitude. In order to probe also regions far from the galactic center, the 382 points are drawn from the N$_{side}$=16 scheme according to the density of star counts. Figure \ref{fig2} illustrates the N$_{side}$=4 basic grid and also the finer grid drawn from the N$_{side}$=16 scheme.

The previously well determined parameters, $\alpha_Y, \alpha_O, z_Y$ and $z_O$ are little affected by the choice of a finer grid. However, $a_H$ and $N_{sph}/N_D$ are. This is due to the fact that the cusp in star counts clearly visible at $-10^\circ < l < +10^\circ$ is well sampled by the finer grid. The trend of the changes is in the sense that $a_H$ falls and $N_{sph}/N_D$ goes up.

\section{Results} \label{section:results}

\subsection{Parameters}
\begin{figure}
\includegraphics[scale=1.0] {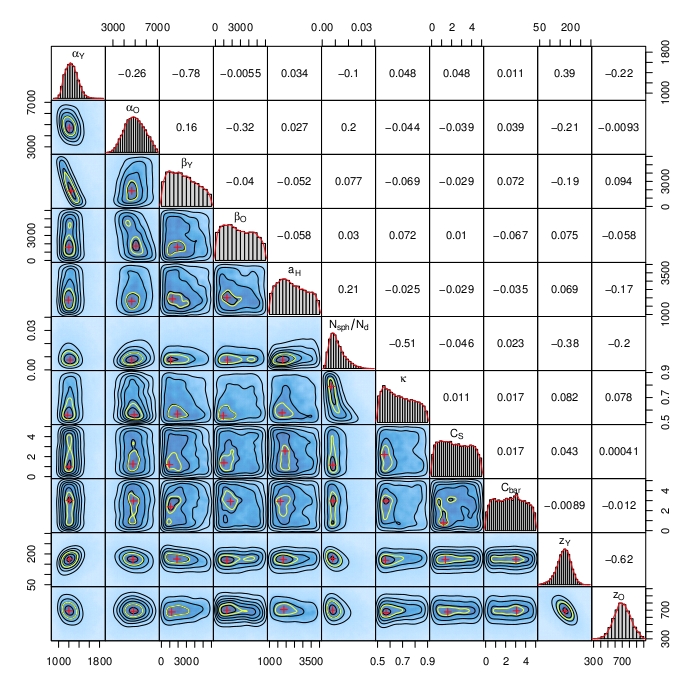}
\caption{Joint \textit{a posteriori} probability densities for the parameters of our model after $10^5$ iterations of a MCMC, considering the N$_{side}$=4 grid. The marginalized 1D histograms for the parameters are displayed on the diagonal. We also included the correlation coefficients between each pair of parameters. The red cross indicates the mode of the 2D distribution and the yellow contour shows the 95\% confidence region.
\label{fig3}}.
\end{figure}

Figure \ref{fig3} shows an overview of the joint distributions of probabilities of the parameters
from a MCMC run of $10^5$ iterations considering the N$_{side}$=4 HEALPix grid. One can see that $\alpha_Y, \alpha_O, z_Y$ and $z_O$ are
the best constrained parameters, followed by $N_{sph}/N_D$ and $\kappa$. 
Two factors contribute for making the
estimates of $a_H$, $\beta_Y$, and $\beta_O$  more difficult: their much more subtle contribution to the
overall behavior of the counts and the relatively poor resolution of the 192 points grid. One
has to recall that the separation of samples in longitude for $b = 0^\circ$ is $22.5^\circ$, and that the first
``parallels'' are at $|b| \sim 9.6^\circ$. The parameter related to the contrast of the spiral arms, $C_S$, and the parameters
of the bar ($C_{bar}, \theta_{bar}, x_0,y_0,z_0$) were kept fixed at the best possible guesses when using the sparse
grid. This means contrasts of the order of the unity, $\theta_{bar} \sim 30^\circ$, $x_0 \sim 2000$ pc and 
$\left\lbrace x_{0}:y_{0}:z_{0}\right\rbrace =  \left\lbrace 1.0:0.4:0.4\right\rbrace$.
 
\begin{figure}
\includegraphics[scale=1.0] {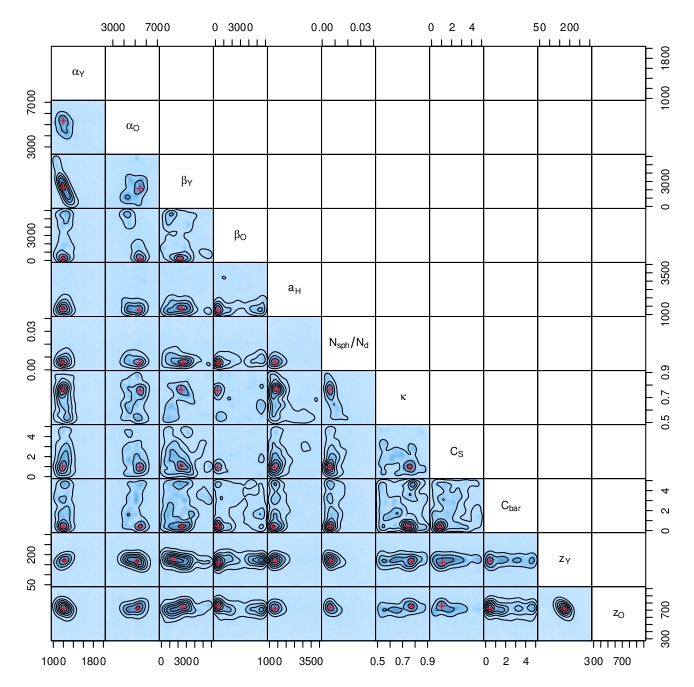}
\caption{The distribution of densities for the parameters of our model after 7400 iterations of a Nested Sampling run with $N_{live}=512$, considering the N$_{side}$=4 grid. The red symbol indicates the mode of the 2D distribution.
\label{fig4}}. 
\end{figure}

\begin{figure}
\includegraphics[scale=1.0] {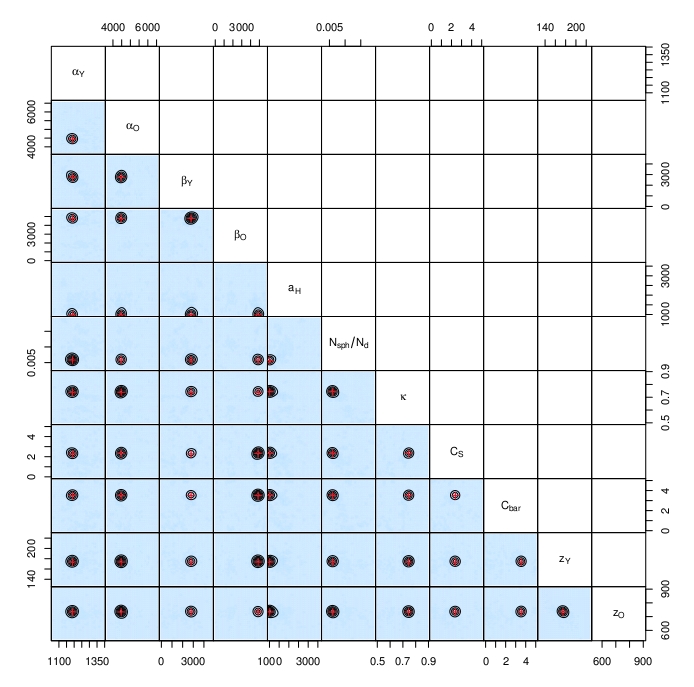}
\caption{The distribution of densities for the parameters of our model from a Nested Sampling run which started with $N_{live}=32$ random states chosen among the ones with the 5\% best likelihoods in a MCMC of 25000 iterations, considering the N$_{side}$=4 grid. The red symbol indicates the mode of the 2D distribution.
\label{fig5}}. 
\end{figure}

Figure \ref{fig4} illustrates our attempt to better constrain the parameters via the NS algorithm.
For $\alpha_Y, \alpha_O, z_Y, z_O$ and $N_{sph}/N_D$ the best solution coincides well with the maximum
likelihoods from the MCMC run. For $\beta_Y, \beta_O, a_H$, $\kappa$, $C_S$ and $C_{bar}$ the maxima tend to fall in regions
that may be far from the correspondent in the MCMC distribution. This just reflects the ``flat'' nature of the likelihood landscape, with small differences between the discrete evaluations which can lead to solutions along a wide range of values. We tried to mitigate this limitation choosing as the start state for the NS algorithm 32 random likelihoods among the 5\% best evaluated in a MCMC run of 25000 steps. The result is shown in Figure \ref{fig5}. We see that the best determined parameters from the MCMC procedure converged to consistent values, while the rest tend -- even though converging to definite values -- to show substantial spread in parameter space as the NS algorithm progressed.

Table\,\ref{tb:tb4} summarizes the results shown in Figs. \ref{fig3}, \ref{fig4} and
\ref{fig5} in numerical form. The well constrained parameters, $\alpha_Y$, $\alpha_O$, $z_Y$ and $z_O$, show consistency in all three approaches and are probably more accurately determined than indicated by the MCMC procedure alone. For the rest of the parameters the range indicated by the MCMC exploration is large and reflects the weak constraints imposed on them by the adopted sampling grid.  

Table\,\ref{tb:tb5} shows the effects of adopting different
uncertainties in the star counts as mentioned in Section 3.1. To make the comparison simple, we chose to minimize only the well determined
parameters. As one can see, the different weighting schemes do not produce conflicting results. In the following,
we discuss only results for which the $\sqrt{N}$ scheme was used.

\begin{figure}
\includegraphics[scale=1.0] {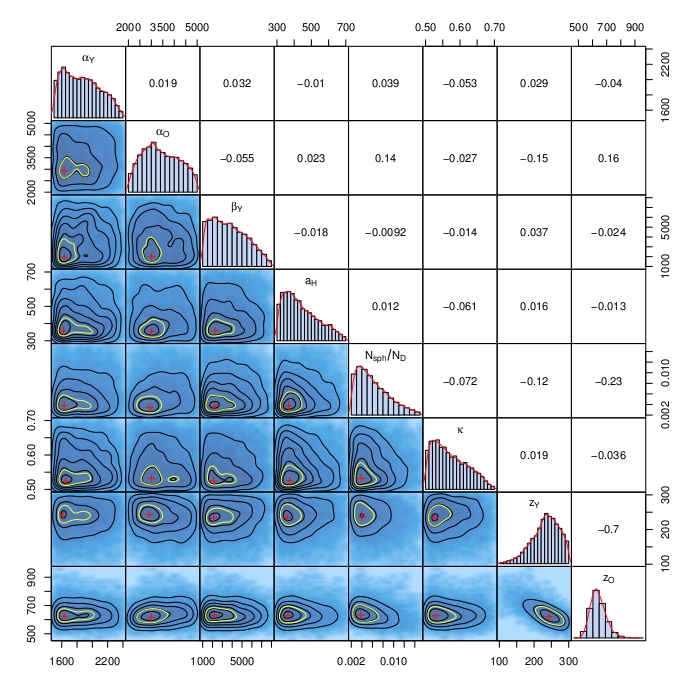}
\caption{Joint \textit{a posteriori} probability densities for the parameters of our model after $7 \times 10^4$ iterations of a MCMC, considering the finer grid of 382 points, shown in Figure \ref{fig2}. The marginalized 1D histograms for the parameters are displayed on the diagonal. We also included the correlation coefficients between each pair of parameters. The symbols are as in Fig. \ref{fig3}.
\label{fig6}}.
\end{figure}

The adoption of a finer grid (see Section 3.3) improves our ability in determining parameters that are related to spatially limited regions of the Galaxy. This is the case for the spheroid, for example. Figure \ref{fig6} shows the result of a MCMC on such a finer grid. Notice that we fixed ill-determined parameters like $\beta_O$, $C_S$ and $C_{bar}$ to the best possible guesses. A further NS run on the region constrained by the MCMC run of Figure \ref{fig6} using the full resolution of a N$_{side}$ = 16 (3072 points) grid, provided us with what we consider the best estimate for the basic galactic parameters from the 2MASS data. They are discussed in the following section.

\subsection{Comparison with results in the literature}

Table \ref{tb:tb6} shows the results of this work together with a compilation of corresponding values found in the literature, to facilitate a comparison and discussion of possible differences. The first result that catches our attention is $\alpha_Y$, for which we obtain values which are systematically smaller than those quoted in the literature, specially considering the results from the N$_{side}=4$ coarse grid. We attribute at least part of the difference to the interplay between $\alpha_Y$ and $\beta_Y$ -- the latter in general not used in models by other authors. $\alpha_O$ is comparable with the estimates of \citet{larsen2003} and \citet{chang}. Like in most cases seen in the literature, we found that $\alpha_{O}$ is larger than $\alpha_{Y}$. \citet{lopezcorredoira2002} present parameters for the galactic disk also based on 2MASS data. They found 2.1 kpc for the scale length of the disk, which is in agreement with what we found for the thin disk.  The value we obtain for $\beta_Y$ is consistent with the estimates by \citet{freuden1998}, \citet{leroy2000}, \citet{lopezcorredoira2004} and \citet{picaud2004}.

The two scale heights, $z_Y$ and $z_O$, are also well determined parameters. $z_Y$ is in good agreement with the values obtained by \citet{robin1996} and 
\citet{juric2008} but slightly smaller than those found by \citet{lopezcorredoira2002}, \citet{reidmaj1993}, and \citet{chang}. $z_O$ is definitively smaller than the values in the literature (\citet{robin1996} , \citet{larsen2003}, \citet{juric2008} and \citet{chang}). \citet{lopezcorredoira2002} determined the scale height of the sole disk in their model equal to 310 pc, which is intermediate between the values  we found in our two-disks model. In this context, it is important to recall that in our model $z_Y$ and $z_O$ follow Eq. \ref{zxr}, and as consequence, any comparison with the literature should refer to the solar neighborhood values.

The scale length parameter of the spheroid, $a_H$, defines how well the ``cusp" in star counts close to the galactic center is fitted. This cusp is conspicuous if we examine the $K_{S}$ longitudinal counts at $|b| \sim 2.4^ \circ$, the first ``parallel" in the N$_{side}$=16 HEALPix scheme. To be able to reproduce the cusp with the spheroidal population of Eq.\,\ref{spheroid}, $a_H$ has necessarily to be small and the normalization $N_{sph}/N_D$ large. The largest values needed for $N_{sph}/N_D$, however, do not
exceed 0.01. The oblateness of the spheroid, $\kappa$, estimated to be $\sim 0.57$, is consistent with the range of values in the literature, 0.55--0.8. We notice that since there is a correlation between this parameter and  $N_{sph}/N_D$, both parameters should always be optimized simultaneously. \citet{larsen2003} suggest that this parameter could vary with galactocentric radius. \citet{carollo2007} concluded that the spheroid would be better described by two subpopulations, one related to an inner bulge, with oblateness $\sim$ 0.6, and another related to an outer bulge, with oblateness $\sim$ 0.9.

A comparison of our results for the spheroid with those of \citet{robin2012} would be interesting since this is a recent result also based on 2MASS data. Unfortunately, it is not straightforward.  Those authors use additional components in their boxy bar/bulge + thicker ellipsoid structure. Besides, they express their results as maps of residuals given by $(C_M - C_{obs})/C_{obs}$ for a very fine grid with 15 arcmin separation in the region $|l| < 20^{\circ}$ and $|b| < 10^{\circ}$. One can see from their Figure 3 that only the central region, a few square degrees in size,  presents residuals well in excess of 20\%. However, their best fit model (case S + E in the paper) involves 15 parameters, while our description uses 7 parameters for the spheroid which are optimized {\em simultaneously} with the parameters that describe the rest of the Galaxy.  Considering the same $20^{\circ} \times 10^{\circ}$ region, our model presents a smooth trend of overestimating the star counts for $|l| \lesssim 10^\circ$, with typical values of the residuals of $\sim 20 \%$. The largest residual in the  $20^{\circ} \times 10^{\circ}$ region is 0.46.

In models where the bulge follows a truncated power law (\citet{binney1997} and \citet{vanholle2009}), the scale length of the bulge plays a different role with respect to our $a_H$ -- it indicates, roughly, the truncation radius of the bar. For this reason we can not quantitatively compare these scales. Similar difficulties are found in the case of the boxy-bulge of \citet{lopezcorredoira2005}. The HWHM of a Gaussian density profile would be 425 pc,  close to our length scale, but the contribution of the boxy-bulge structure extends to larger galactocentric radii.

The parameter describing the contrast of the spiral arms, $C_S$, is very loosely constrained by the 2MASS data, in fact, even with the N$_{side}=16$ grid which gives a separation of $5.6^\circ$ at $|b|=0^\circ$, the enhancement of the counts in the tangential directions mentioned at the end of section 2.3 are hardly seen. Clearly, higher resolution is needed to characterize this component. A similar conclusion was reached by \citet{quillen2002}. We find indications that $C_S$ is of the order of unity, roughly consistent with previous estimates from \citet{drimmel2001}, \citet{grosbol2004}, \citet{benjamin2005} and \citet{liu2012}.

The inclusion of a bar in the modeling of the data sampled with the N$_{side}=16$ HEALPix scheme definitely improves the quality of the resulting fits, even though the limit of  K$_S=11$ and the spacing of $\Delta l = 5.6^\circ$ and $\Delta b = 2.4^\circ$ not being the best to constrain that feature. We find that the bar has a half length of  $\approx 1.25$ kpc, with axes ratios 1.00:0.22:0.39. The angle $\theta_{bar}$ is close to the lower limit of the values found in the literature, $\sim 12^\circ$ \citep{lopezcorredoira2000}. The contrast of the bar with the ambient disks, $C_{bar}$, is $\sim 3$. We did not find previous estimates of the latter quantity in the literature.

\subsubsection{Comparison in selected lines-of-sight}

\begin{figure}
 \includegraphics[scale=0.65,angle=270] {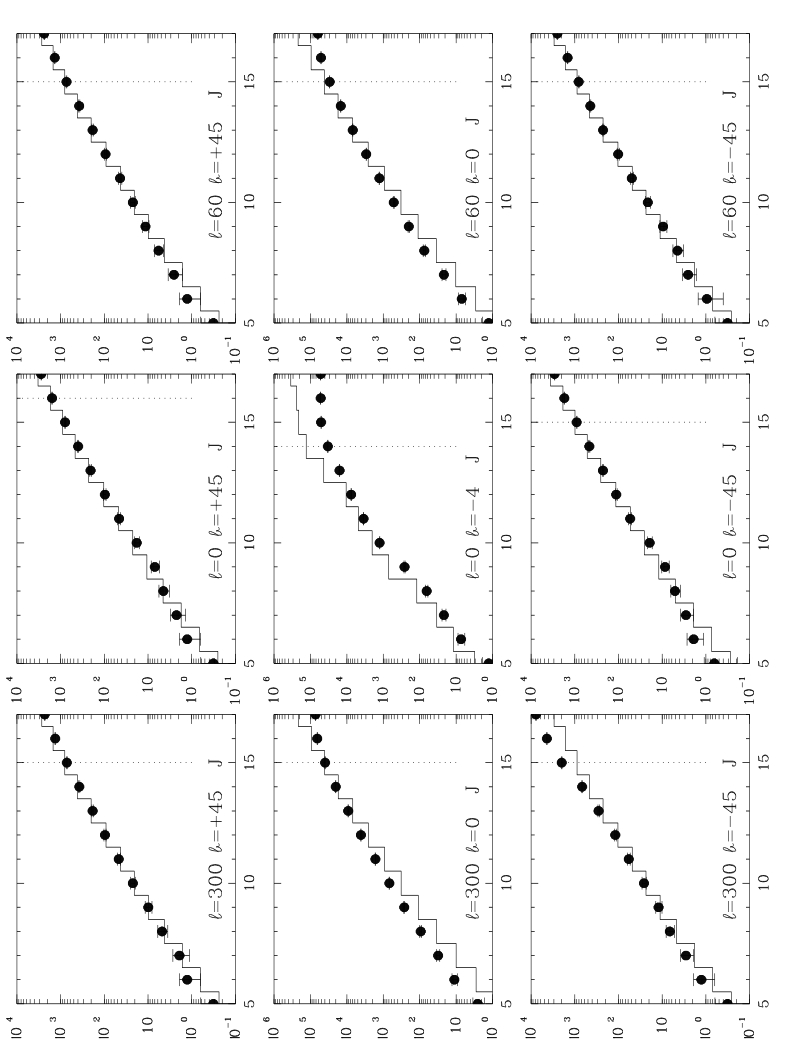}
 \caption{Cumulative star counts histograms for selected lines-of-sight in the J band. The filled circles with error bars represent the observations while the steps 
  correspond to the model. The three middle histograms refer to the galactic plane or close to it, while the upper and lower histograms correspond to  $|b| = 45^\circ$. The dashed line indicates the magnitude limit for each line-of-sight.
 \label{fig7}}. 
 \end{figure}

\begin{figure}
 \includegraphics[scale=0.65,angle=270] {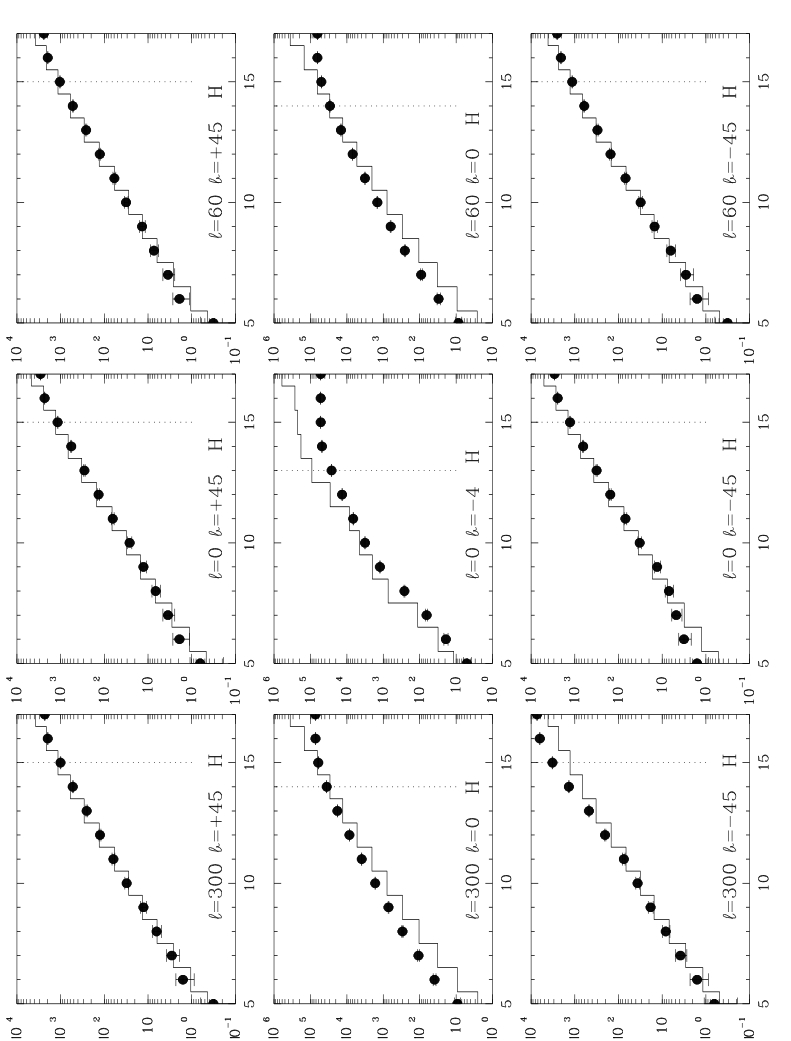}
 \caption{Same as Figure \ref{fig7} for the H band. \label{fig8}}. 
 \end{figure}

\begin{figure}
 \includegraphics[scale=0.65,angle=270] {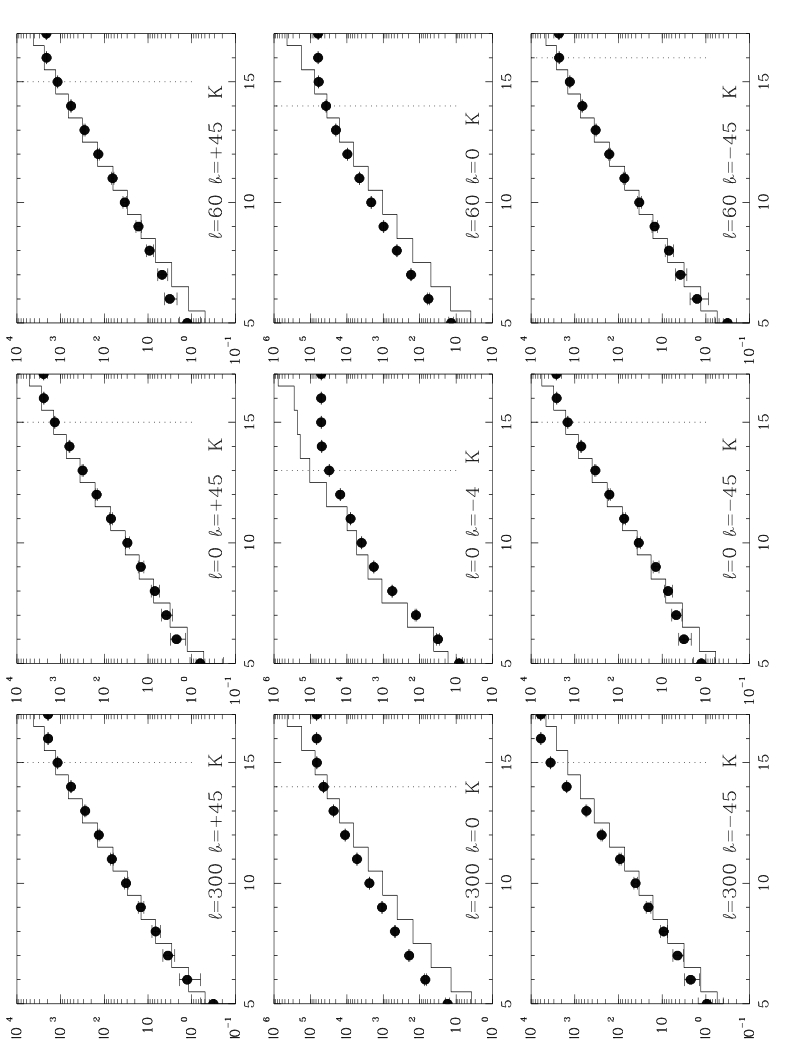}
 \caption{Same as Fig. \ref{fig7} for the K$_{S}$ band. \label{fig9}}. 
 \end{figure}

\begin{figure}
 \includegraphics[scale=0.65,angle=270] {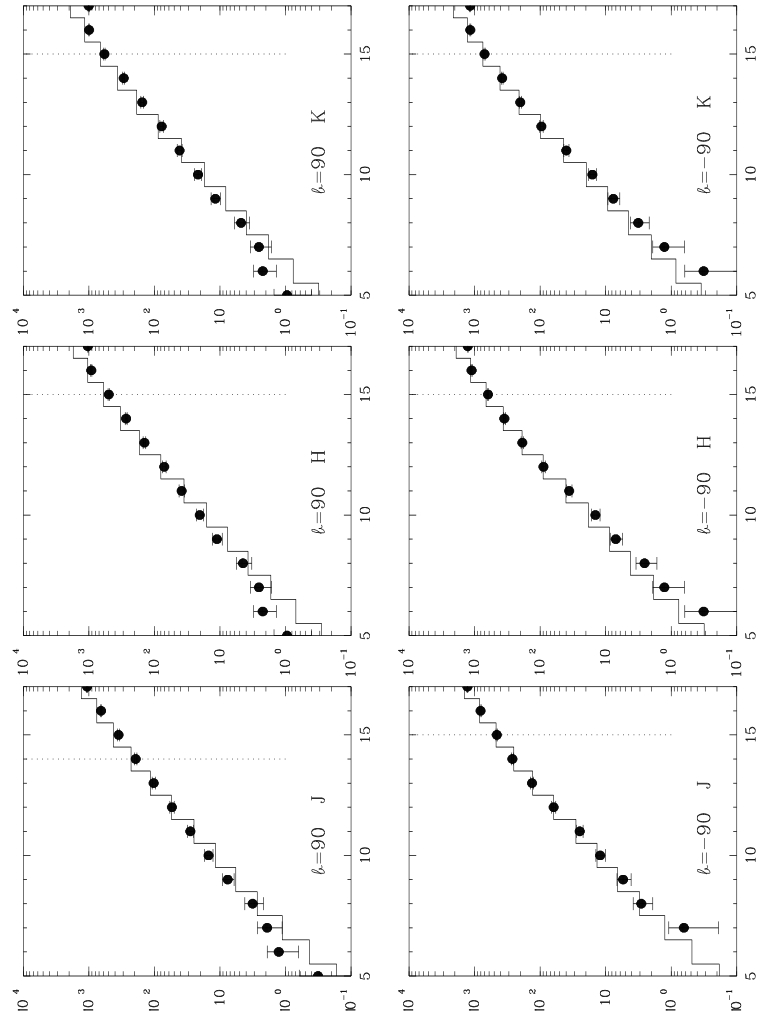}
\caption{Cumulative star counts histograms for the galactic poles in JHK$_{S}$ bands. Symbols are the same as in Fig. \ref{fig7}. \label{fig10}}
 \end{figure}

Figures \ref{fig7}, \ref{fig8} and \ref{fig9} show cumulative histograms of star counts in nine sets of 
$\{l,b\}$ in JHK$_{S}$, respectively. We see that the largest
differences happen close to the galactic plane, specially for $l \sim 300^\circ$ and $l \sim 60^\circ$. One can easily see the effects of the proximity to the Small Magellanic Cloud in the line-of-sight corresponding to $(l,b) = (300,-45)$ for magnitudes fainter than $\approx 13$. Figure \ref{fig10} shows  similar histograms   for $|b| = 90^\circ$. Here we see that the limit magnitudes go all the way to the depth of the 2MASS catalog.

\begin{figure}
\includegraphics[scale=0.8,angle=0] {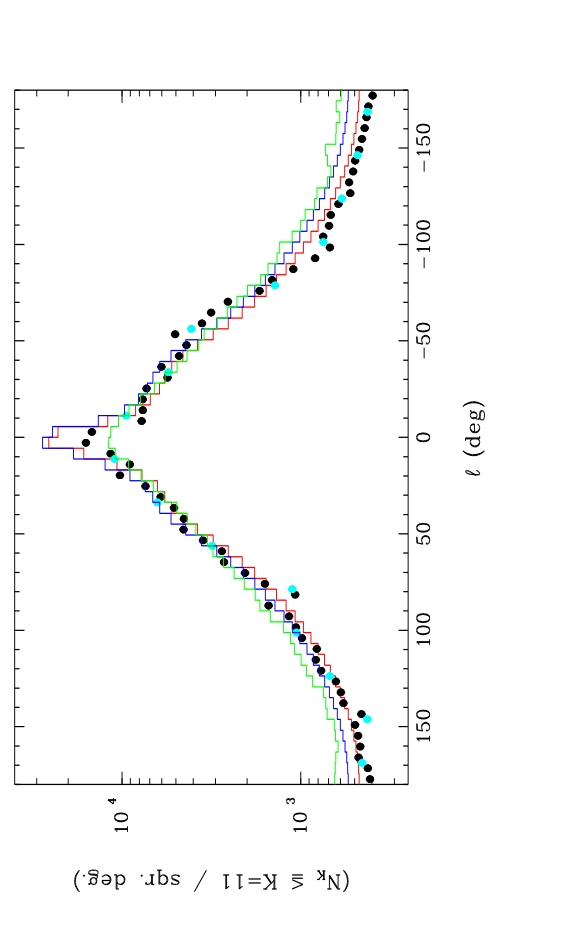}
\caption{Cumulative star counts along longitude for $b=0^{\circ}$ in the K$_{S}$ band. The black/cyan filled circles with error bars represent the observations for the N$_{side}$=4/16 HEALPix grid. The blue and red steps correspond to the model with and without the
correction of Eq. \ref{densjacques}
for thin disk density, respectively. The green steps indicate the
counts calculated with the Besan\c{c}on model. \label{fig11}}.
\end{figure}

\begin{figure}
 \includegraphics[scale=0.8,angle=0] {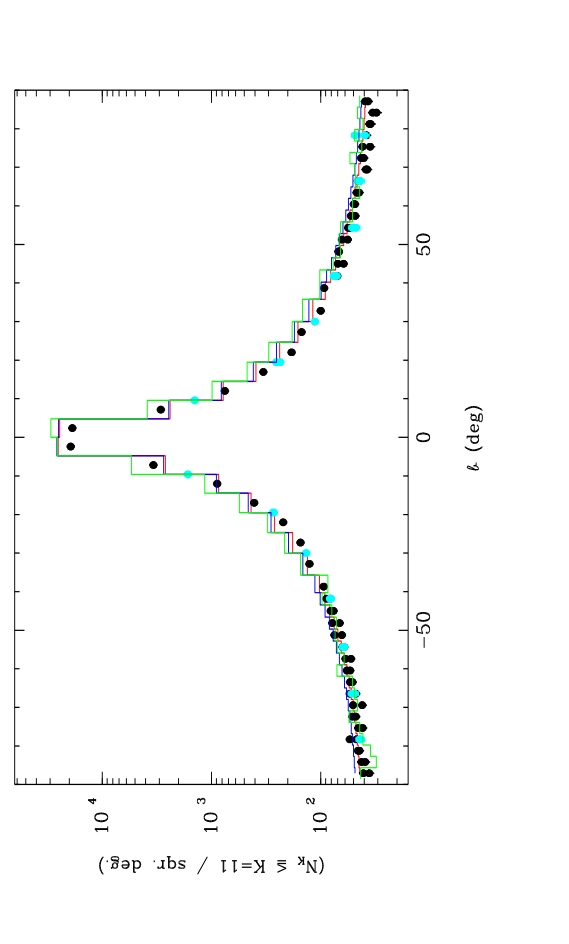}
  \caption{Cumulative star counts along latitude for $l=0^{\circ}$ in the K$_{S}$ band.
 The symbols are as in  Fig. \ref{fig10}, with green steps indicating the counts calculated with the Besan\c{c}on model.   \label{fig12}}. 
 \end{figure}

Counts along the galactic plane, or intercepting the same, use to be a {\em tour de force} for star counts models. Figure \ref{fig11} shows the comparison between observations and model for 64 lines-of-sight in the galactic plane, with and without the adoption of the density profile for the young disk given by Equation \ref{densjacques}. The limit magnitude is 11 in the $K_{S}$ band. For comparison, we also show the predicted counts from the Besan\c{c}on model \citep{besancon,robin2003}, obtained from the online 
form\footnote{http://model.obs-besancon.fr/}. Since the only parameter which can be varied by the user in the online form is the extinction, we show the fit which provides the best description close to the galactic center. Figure \ref{fig12} shows the observed and predicted counts along galactic latitude for $l=0^\circ$.  The largest differences are confined to the region close to the galactic plane. Again, for comparison, we show the predicted counts from the Besan\c{c}on model. Comparing our counts to those from the Besan\c{c}on model, one can see that our fit is slightly better.

\begin{figure}
\centering
{\includegraphics[scale=0.4,angle=180.0]{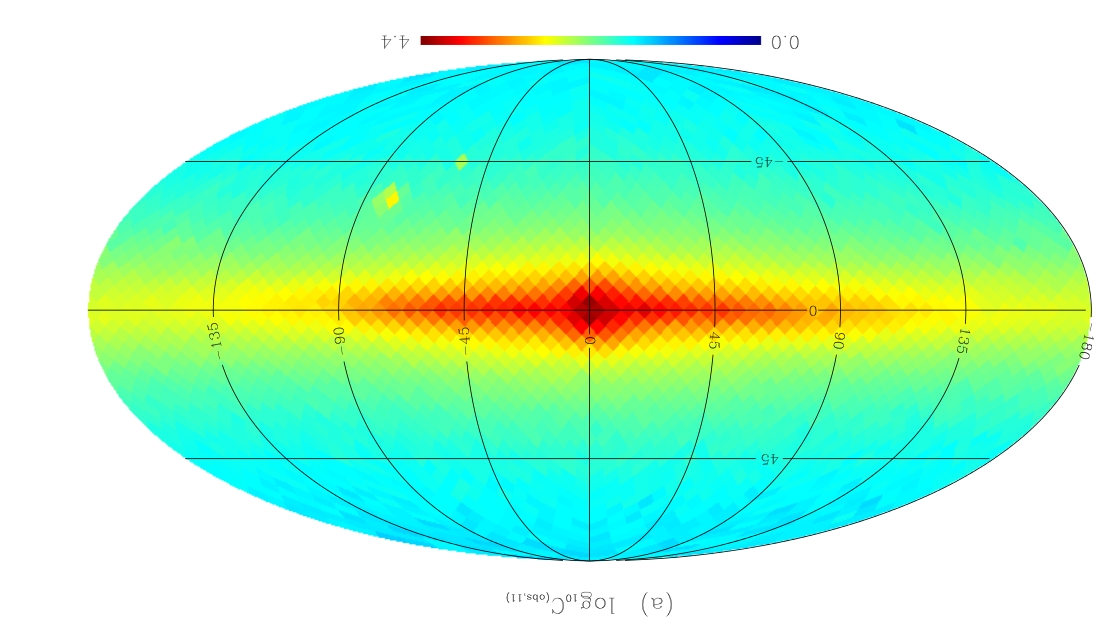}}
{\includegraphics[scale=0.4,angle=180.0]{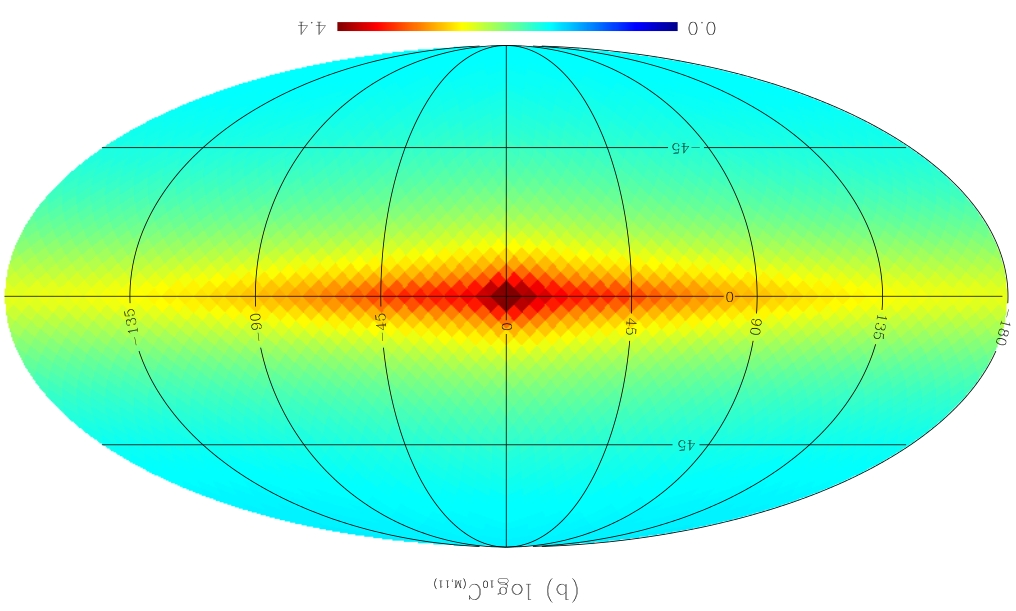}}\\
{\includegraphics[scale=0.4,angle=180.0]{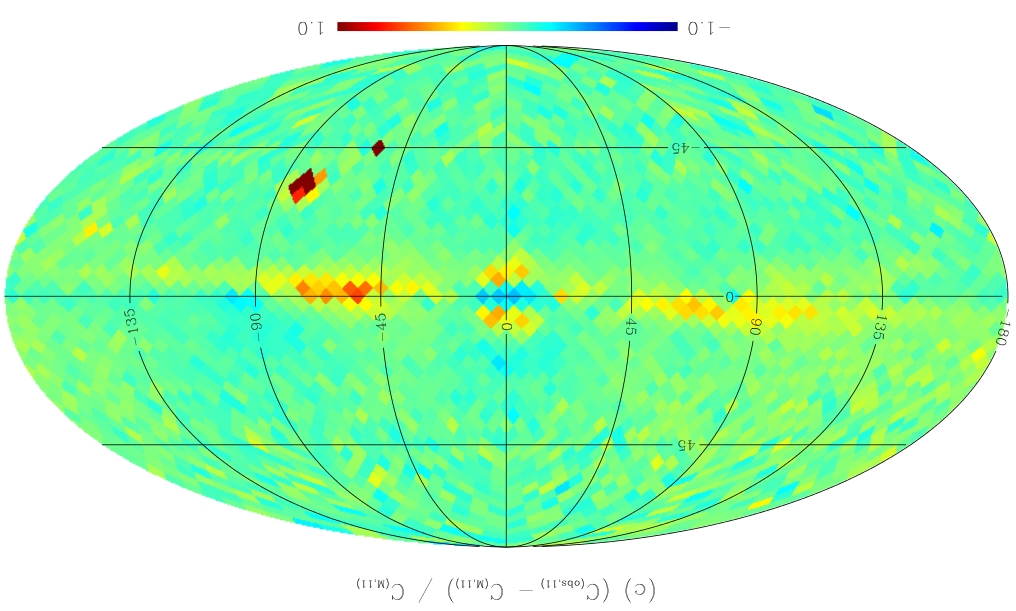}}\\
\caption{\textbf{(a)} The observed cumulative star counts for K$_S < 11$ sampled according to the N$_{side}$=16 HEALPix scheme (3072 grid points). Each grid point is the result of a cone search of one square degree area. The counts are color-coded in a logarithmic scale to facilitate visualization. \textbf{(b)} The predicted cumulative star counts from our model in the same band with the same counts coding as in (a). \textbf{(c)} The relative differences $(C_{(obs,11)}-C_{(M,11)})/C_{(M,11)}$ of (a) and (b)  color-coded in a
 {\em linear scale}, to emphasize the details.}
\label{fig13}
\end{figure}

An all sky map of the relative differences between observed and predicted counts produces a comprehensive view of the merits and limitations of a model. 
Fig.\,\ref{fig13}(c)  shows as a figure of merit the ratio 
$(C_{obs}-C_{M})/C_{M}$ obtained from the cumulative histograms up to magnitude 11. This is not very different but a little more intuitive than the figure of merit  ${C}_{M}/{C}_{obs}$ used by \citet{chang}. For reference, both the observed and model counts are presented
in Figs.\,\ref{fig13}(a) and \ref{fig13}(b), respectively. As we can see in Fig.\,\ref{fig13}(c), the largest differences between observed and predicted counts are concentrated in the galactic plane. This is a limitation shared by all models in the literature.
Our Fig.\,\ref{fig13}(c) resembles  Figure 2 of \citet{reyle2009}, both presenting the differences between observations and model predictions. Although the model of \citet{reyle2009} includes  a description of a warp, not present in our model, the O-C differences are smaller in our case.
The main differences between our map and that one from \citet{reyle2009} are the locations where the galactic model does not coincide with observations. The excess of observed stars counts which they attribute to the warp  is present in the outer Galaxy, while the largest excesses observed by us are close to $l \sim 300^\circ$ and $l \sim 60^\circ$.

\begin{figure}
 \includegraphics[angle=180.,scale=0.70] {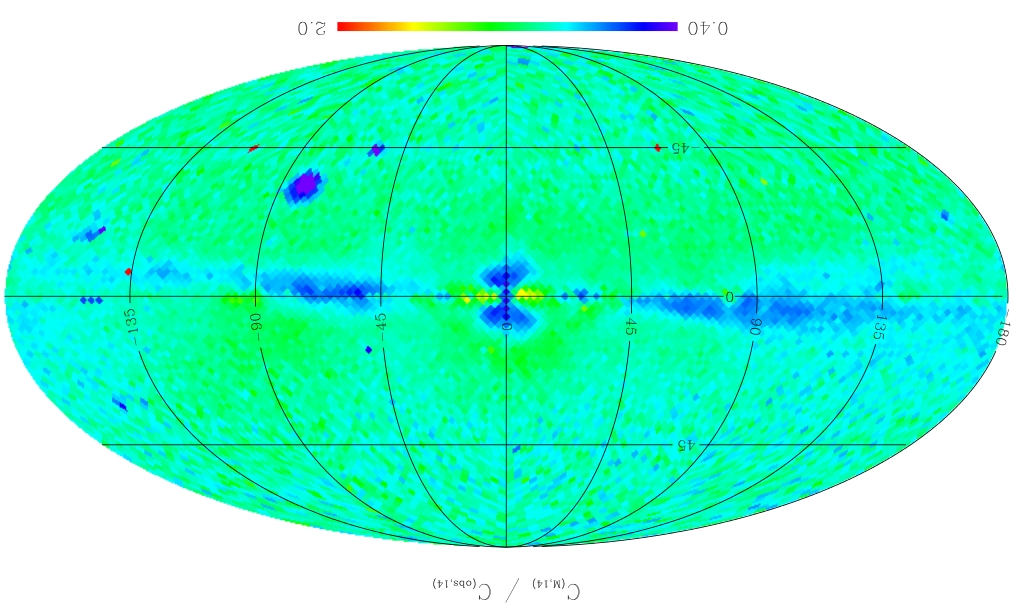}
 \caption{The ratio $C_{(M,14)}/C_{(obs,14)}$ in the K$_S$ band for a 12288 grid points from the N$_{side}=32$ HEALPix scheme, for a direct comparison with the results of \citet{chang}. The cumulative limiting magnitude
 in this case is 14.   \label{fig14}}. 
 \end{figure}

Fig.\,\ref{fig14} shows an all sky map of ${C}_{M}/{C}_{obs}$ for a more direct comparison with the result obtained by \citet{chang}. Our description is better both at the center and anti-center regions.

\section{Conclusions}

We have used the 2MASS data to test a modified version of the model due to OL93 in the JHK$_S$ bands. We emphasize that this study is the first attempt to determine the main parameters describing the Milky Way using the observed star counts in the whole sky, including the galactic plane.

A survey of the parameter space was done using the Markov chain Monte Carlo method, followed by attempts to better constrain the parameters via the Nested Sampling algorithm. A grid with 192 points generated by the HEALPix scheme was adopted as a baseline for sampling and provided good initial estimates for $\alpha_{Y}$, $\alpha_O$, $z_Y$, $z_O$, $N_{sph}/N_D$ and $\kappa$. Since we expected poor results from the coarse grid considering that regions close to the galactic center  and galactic plane show large gradients in star counts, finer grids were built to better sample those regions.  In general, our results show a good agreement with the values of $\alpha_{Y}$, $\alpha_{O}$, $z_Y$, $z_O$, $N_{sph}/N_D$ and $\kappa$ found in the literature. We find that the ``hole in the disk" ($\beta$ radial component) shows a strong anti-correlation with $\alpha$. The value of $a_{H}$ found in this study describes the cusp in star counts close to the galactic center and 
this is the reason why it differs in scale from determinations of size for the spheroidal component found in the literature which describe larger scales.
We find a moderate anti-correlation between the oblateness of the spheroid, $\kappa$, and the constant of normalization, $N_{sph}/N_D$. This is important and may happen in other models too. One has to have in mind that both parameters should be optimized simultaneously. There is also a definite anti-correlation between $z_Y$ and $z_O$. These parameters, however, are very well defined, even with the use of a coarse grid.

Our model describes the star counts in 80\% of the sky with an accuracy of better than 10\%. In the remaining area, a few dozen lines-of-sight (out of 3072) show absolute residuals in excess of 20\%. They are concentrated close to the plane ($|b| <$ 8$^\circ$), specially around $l \sim 300^\circ$, $l \sim 60^\circ$ and surrounding the galactic center.

An overall view of the Galaxy according to our work suggests that it can be described by two disks with radial scales slightly shorter than those found in the literature. Only the thick disk needs a ``hole" in its inner part. The conspicuous concentration of sources in the direction of the galactic center can be described by a combination of contributions from the disks, from a spheroid with scale 400 pc, and from a bar that has aspect ratios 1.00:0.22:0.39, and is seen at an angle of 12$^{\circ}$.

\section{Acknowledgments}

PFP received financial support for this work from \textit{Coordena\c{c}\~ao de Aperfei\c{c}oamento de Pessoal de N\'ivel Superior}. This publication makes use of data products from the Two Micron All Sky Survey, which is a joint project of the University of Massachusetts and the Infrared Processing and Analysis Center/California Institute of Technology, funded by the National Aeronautics and Space Administration and the National Science Foundation.

\clearpage



 \clearpage


\clearpage




\clearpage

\clearpage

\clearpage


\clearpage


\clearpage
%
 \clearpage
 %
 \clearpage
%
 \clearpage

\clearpage

 \clearpage


\clearpage

 \clearpage








\clearpage
\begin{deluxetable}{ccl}
\tabletypesize{\scriptsize}
\tablecaption{Counting rejections based on quality flags.\label{tb:tb1}}
\tablewidth{0pt}
\tablehead{
\colhead{Flag} & \colhead{Catalog Code} & \colhead{Meaning}
}
\startdata

Photometric quality  &  X,U\tablenotemark{a}   &   Objects not detected in one of the bands \\
 &   & or from which it was not possible \\
 &   & to estimate the brightness \\

Read  &  0, 9\tablenotemark{b}   &  Objects not detected in one of the bands \\
 &   & or, although detected, it was not possible\\
 &   & to make useful estimates of brightness \\

Contamination and confusion   &   p, c, d, s, b\tablenotemark{c}   &  Objects affected by the proximity of \\
 &   &  a bright source or image artifacts   \\

Galactic contamination  &   1, 2\tablenotemark{d}   &   Objects contaminated by the proximity of \\
 &   &  extended sources  \\

\enddata
\tablenotetext{a}{Possible codes for this flag are X, U, F, E, A, B, C, D}
\tablenotetext{b}{Possible codes for this flag are 0, 1, 2, 3, 4, 6, 9}
\tablenotetext{c}{Possible codes for this flag are p, c, d, s, b, 0}
\tablenotetext{d}{Possible codes for this flag are 0, 1, 2}
\end{deluxetable}

\clearpage
\begin{deluxetable}{ccccccccccc}
\tabletypesize{\scriptsize}
\tablecaption{Percentage of rejections for selected lines-of-sight in J, H and K$_{S}$.\label{tb:tb2}}
\tablewidth{0pt}
\tablehead{
\colhead{($l$,$b$)} & \colhead{Band} & \colhead{$m=9.5$} & \colhead{$m=10.5$} & \colhead{$m=11.5$} & \colhead{$m=12.5$} & \colhead{$m=13.5$} 
}
\startdata
(0,+45)    &   J   &  0.0  & 0.0   &  0.5 & 1.6 & 2.9  \\
           &   H   & 0.2 & 0.2 & 1.1 & 2.6 & 3.4  \\
           &   K$_{S}$   & 0.1 & 0.0  & 0.2 & 0.3 & 1.8  \\

(60,+45)   &  J   & 0.2 & 0.2 & 0.2 & 1.2 & 7.5  \\
           &  H   & 0.4 & 0.6 & 0.2 & 2.6 & 9.4  \\
           &  K$_{S}$  & 0.2 & 0.1 & 0.3 & 1.6 & 3.8  \\

(300,+45)  &  J    & 0.7 & 0.7 & 0.3 & 2.0 & 3.3  \\
           &  H  & 0.4 & 0.4 & 0.2 & 1.3 & 5.7  \\
           &  K$_{S}$   & 0.2 & 0.1 & 0.2 & 0.7 & 2.2  \\

(0,-4)    &  J   & 0.3 & 1.0 & 2.7 & 8.8 & 33.2  \\
          &  H   & 1.0 & 2.4 & 6.8 & 22.0 & 53.7  \\
          &  K$_{S}$  & 1.2 & 3.0 & 8.2 & 26.9 & 40.0  \\

(60,0)    &  J  & 0.0 & 0.1 & 0.5 & 1.6 & 5.0  \\
          &  H  & 0.2 & 0.6 & 1.8 & 5.0 & 15.2  \\
          &  K$_{S}$  & 0.3 & 0.9 & 2.7 & 6.0 & 21.8  \\

(300,0)   &  J  & 0.0 & 0.1 & 0.5 & 1.7 & 5.9  \\
          &  H  & 0.2 & 0.5 & 1.6 & 5.2 & 16.0  \\
          &  K$_{S}$  & 0.3 & 0.9 & 2.8 & 8.2 & 23.7  \\

(0,-45)   &  J   & 0.0 & 0.4 & 0.2 & 1.1 & 4.4  \\
          &  H   & 0.3 & 0.3 & 0.3 & 3.0 & 5.0  \\
          &  K$_{S}$  & 0.1 & 0.1 & 0.4 & 1.4 & 3.0  \\

(60,-45)  &  J   & 0.3 & 0.3 & 0.3 & 1.1 & 3.8  \\
          &  H   & 0.2 & 0.2 & 0.2 & 1.9 & 4.2  \\
          &  K$_{S}$   & 0.0 & 0.2 & 0.2 & 1.0 & 2.9  \\

(300,-45) &  J   & 0.1 & 0.1 & 0.3 & 0.5 & 3.0  \\
          &  H  & 0.1 & 0.1 & 0.3 & 0.7 & 3.9  \\
          &  K$_{S}$   & 0.0 & 0.1 & 0.1 & 0.5 & 2.4  \\
\enddata
\end{deluxetable}


\clearpage
\begin{deluxetable}{llcc}
\tabletypesize{\scriptsize}
\tablecaption{Limits for parameters search.\label{tb:tb3}}
\tablewidth{0pt}
\tablehead{
\colhead{Parameter} & Symbol & \colhead{Lower} & \colhead{Upper}
}
\startdata
Radial scale length of thin/thick disks (pc)        & $\alpha_Y,\alpha_O$   &  500   &   7000    \\
Radii of the central hole in thin/thick disks (pc)  & $\beta_Y, \beta_O$     &    0   &   6000    \\
Length parameter of the spheroid (pc)                & $a_H$                          & 100   &   4000    \\
Spheroid to disk density ratio                             & $N_{sph}/N_{D}$              &   0.001   &    0.020     \\
Oblate spheroid parameter                                 & $\kappa$                     &    0.5    &     0.9      \\
Density contrast of the spiral arms                     & $C_S$                          &    0.0    &    3.0      \\
Density contrast of the bar                                 & $C_{bar}$                     &    0.0    &    5.0      \\
Scale height of the thin disk (pc)                         & $z_Y$   &    50  &   400     \\
Scale height of the thick disk (pc)                       & $z_O$   &   200  &  1000     \\ 
Bar half length (pc)                                           &   \textit{l}$_{bar}$      &   700     &    4000   \\
Orientation angle of the bar (deg)                        & $\theta_{bar}$  &  11      &     53      \\
\enddata
\end{deluxetable}


\clearpage
\begin{deluxetable}{cccc}
\tabletypesize{\scriptsize}
\tablecaption{Results from MCMC, NS and MCMC+NS for the N$_{side}$=4 HEALPix grid.\label{tb:tb4}}
\tablewidth{0pt}
\tablehead{
\colhead{Parameter} & \colhead{MCMC} & \colhead{NS} &  \colhead{MCMC+NS} }
\startdata

$\alpha_Y$ (pc)    & $\left(1230_{-170}^{+190}\right)$ &  $\left(1200_{-170}^{+190}\right)$  & $\left(1190_{-160}^{+170}\right)$ \\
$\alpha_O$ (pc)    & $\left(4750_{-690}^{+920}\right)$  &  $\left(5150_{-690}^{+920}\right)$   & $\left(4420_{-750}^{+850}\right)$  \\ 
$\beta_Y$ (pc) & $\left(920_{-570}^{+2570}\right)$ &  $\left(2140_{-570}^{+2570}\right)$                 & $\left(2770_{-600}^{+1700}\right)$  \\ 
$\beta_O$ (pc) & $\left(1740_{-1430}^{+2570}\right)$ &  $\left(100_{-100}^{+2570}\right)$   & $\left(4760_{-730}^{+1240}\right)$   \\
$a_H$ (pc)             & $\left(1940_{-640}^{+570}\right)$ &  $\left(1350_{-640}^{+570}\right)$  & $\left(1000_{-410}^{+550}\right)$  \\
$N_{sph}/N_{D}$              & $\left(0.0073_{-0.0036}^{+0.0064}\right)$ &  $\left(0.0058_{-0.0036}^{+0.0064}\right)$   & $\left(0.0058_{-0.0033}^{+0.0072}\right)$  \\
$\kappa$                & $\left(0.55_{-0.05}^{+0.10}\right)$ &  $\left(0.76_{-0.05}^{+0.10}\right)$   & $\left(0.74_{-0.04}^{+0.09}\right)$  \\ 
$C_S$  &   $\left(0.9_{-0.5}^{+3.3}\right)$  &   $\left(1.0_{-0.5}^{+3.3}\right)$ &    $\left(2.3_{-0.7}^{+1.1}\right)$         \\ 
$C_{bar}$     &   $\left(3.0_{-2.5}^{+0.9}\right)$ &  $\left(0.4_{-0.4}^{+0.9}\right)$ &   $\left(3.5_{-0.3}^{+1.5}\right)$   \\
$z_Y$ (pc)             &  $\left(170_{-30}^{+40}\right)$ &  $\left(170_{-30}^{+40}\right)$  & $\left(170_{-40}^{+40}\right)$   \\ 
$z_O$ (pc)            &  $\left(680_{-90}^{+120}\right)$ &  $\left(710_{-90}^{+120}\right)$  & $\left(730_{-90}^{+130}\right)$  \\ 
\textit{l}$_{bar}$\tablenotemark{*} (pc)     &   2000     &    2000  &   2000   \\
$\theta_{bar}$\tablenotemark{*} (deg)        &  30       &     30   &  30    \\

\enddata
\tablenotetext{*}{These parameters were kept fixed at the values indicated.}
\end{deluxetable}


\clearpage
\begin{deluxetable}{ccc}
\tabletypesize{\scriptsize}
\tablecaption{Comparison between different weighting schemes.\label{tb:tb5}}
\tablewidth{0pt}
\tablehead{
\colhead{Parameter} & \colhead{$\sqrt{N}$} & \colhead{5$-$points variance}}
\startdata
$\alpha_Y$ (pc) & $\left(1200_{-140}^{+220}\right)$  & $\left(1330_{-190}^{+190}\right)$   \\
$\alpha_O$ (pc) & $\left(5150_{-1090}^{+520}\right)$ & $\left(4970_{-550}^{+730}\right)$  \\ 
$z_Y$ (pc)      & $\left(170_{-30}^{+40}\right)$ & $\left(180_{-30}^{+40}\right)$  \\ 
$z_O$ (pc)      & $\left(710_{-120}^{+90}\right)$ & $\left(770_{-60}^{+130}\right)$  \\ 
\enddata
\end{deluxetable}

%
\clearpage
\begin{deluxetable}{cccc}
\tabletypesize{\scriptsize}
\tablecaption{Comparison with results in the literature.\label{tb:tb6}}
\tablewidth{0pt}
\tablehead{
\colhead{Parameter} & \colhead{Our result} & \colhead{Value from literature}& \colhead{Source}}
\startdata
$\alpha_{Y}$ (pc)   & (2120$\pm$200) & (2500$^{+800}_{-600}$) &  \citet{fuxmartinet} \\
                    &                &      2600      &  \citet{freuden1998} \\
                    &                & (2100$\pm$300) &  \citet{porcel1998}  \\
                    &                & (3300$\pm$600) &  \citet{feast2000}   \\
                    &                &       1700     &  \citet{leroy2000}   \\
                    &                & (2800$\pm$300) &  \citet{ojha2001}    \\
                    &                &  $\left(2100_{-170}^{+220}\right)$ &  \citet{lopezcorredoira2002}    \\
                    &                & (3500$\pm$300) &  \citet{larsen2003}  \\
                    &                &    2400        &  \citet{picaud2004}   \\
                    &                & (2600$\pm$520) &  \citet{juric2008}   \\
                    &                &      2200      &  \citet{reyle2009}   \\
                    &                & (3700$\pm$1000) &  \citet{chang}      \\ 
$\alpha_{O}$ (pc)   & (3050$\pm$500) &     3500      &  \citet{reidmaj1993} \\ 
                    &                & (2800$\pm$800) &  \citet{robin1996}   \\
                    &                & (3000$\pm$1500) &  \citet{buser1999}  \\
                    &                &      2300      &  \citet{leroy2000}   \\ 
                     &                & (3700$\pm$800) &  \citet{ojha2001}    \\
                     &                & (4700$\pm$200) &  \citet{larsen2003}  \\                                              
                     &                & (3600$\pm$720) &  \citet{juric2008}   \\
                     &                & (5000$\pm$1000) &  \citet{chang}      \\ 
 $\beta_Y$ (pc)      & $\left(2070_{-800}^{+2000}\right)$ & 3000   & \citet{freuden1998} \\ 
                     &               & 2600        &  \citet{leroy2000} \\
                     &               & 2000 - 4000   & \citet{lopezcorredoira2004}     \\
                     &               & (1310$\pm$1030)   &  \citet{picaud2004}   \\
 $z_Y$ (pc)          &  (205$\pm$40)  &   325      &  \citet{reidmaj1993} \\ 
                     &               &  250-270   &  \citet{robin1996}   \\   
                     &                &   100     &  \citet{leroy2000}   \\
                     &                &   $\left(310_{-45}^{+60}\right)$    &  \citet{lopezcorredoira2002}   \\
                     &               & (245$\pm$49) &  \citet{juric2008} \\
                     &               & (360$\pm$10) &  \citet{chang}     \\
 $z_O$ (pc)          &  (640$\pm$70) & 1400-1600  & \citet{reidmaj1993}  \\ 
                     &               &  760       &  \citet{robin1996}   \\
                     &               &  390       &  \citet{leroy2000}   \\  
                     &               &  900       &  \citet{larsen2003}  \\
                     &               & (900$\pm$180) & \citet{juric2008} \\
                     &               & (1020$\pm$30) &  \citet{chang}    \\  
 $a_H$ (pc)          & (400$\pm$100) &  3000      &  \citet{gilmore1984} \\
                     &               &  2670      &  \citet{reidmaj1993} \\
                     &               &  1900$^{*,a}$    &  \citet{binney1997} \\
                     &               &  420       &  \citet{leroy2000}   \\
                     &               & (4300$\pm$700) & \citet{larsen2003} \\ 
                     &               &  (2500$^{+1730}_{-160}$)$^{*,b}$    &  \citet{vanholle2009} \\
 $N_{sph}/N_{D}$       &  (0.0082$\pm$0.0030) & 0.00125 &  \citet{bahcallsoneira} \\
                     &                     & 0.0083  &  \citet{gugli1990}     \\
                     &                     & 0.00358 &  \citet{ruelas1991}    \\
                     &                     & (0.002-0.003) &  \citet{larsen2003} \\
                     &                     &    0.0051     &  \citet{juric2008}  \\
                     &                     & (0.002$\pm$0.001) &  \citet{chang}  \\ 
 $\kappa$            & (0.57$\pm$0.05)      & (0.80$\pm$0.05) & \citet{reidmaj1993} \\ 
                     &                   &       0.8     &  \citet{leroy2000}     \\
                     &                   &       0.6       &  \citet{robin2000}   \\
                     &                   & (0.55$\pm$0.06) &  \citet{chen2001}    \\ 
                     &                   & (0.65$\pm$0.05) &  \citet{girardi2005} \\
                     &                   & (0.64$\pm$0.01) &  \citet{juric2008}   \\ 
 $C_S$               &   $\left(2.0_{-0.8}^{+0.6}\right)$  &  1.32               & \citet{drimmel2001}  \\  
                     &                               &  1.2 -- 1.4          & \citet{grosbol2004}        \\
                     &                               &  1.30               & \citet{benjamin2005}       \\
                     &                               &  1.3 -- 1.5               & \citet{liu2012}       \\
 $C_{bar}$           &   $\left(3.4_{-1.5}^{+1.0}\right)$ &                 &            \\
 $l_{bar}$ (pc)          &  $\left(1250_{-250}^{+500}\right)$ &    1610 -- 2030    &   \citet{dwek1995}         \\
                         &        &     900    &   \citet{stanek1997}         \\
                         &        &  $<$ 3128    &   \citet{freuden1998}         \\
                         &        &    1750    &   \citet{bissantz2002}         \\
                         &        &    3900    &   \citet{lopezcorredoira2007}         \\
                         &        &    $\sim$ 1250    &   \citet{gonzalez2011}         \\
                         &        &    $\sim$ 1460    &   \citet{robin2012}         \\
                         &        &    $\sim$ 1490    &   \citet{wang2012}         \\
                         &        &    $\sim$ 680    &   \citet{cao2013}         \\
 $\theta_{bar}$ (deg)    &  $\left(12_{-1}^{+15}\right)$    &    (20$\pm$10)    &     \citet{dwek1995}       \\
                         &        &    20 -- 30        &   \citet{stanek1997}         \\
                         &        &    $\sim$ 14   &   \citet{freuden1998}         \\
                         &        &    12        &   \citet{lopezcorredoira2000}         \\
                         &        &    15 -- 30        &   \citet{bissantz2002}         \\
                         &        &    20 -- 35        &   \citet{lopezcorredoira2005}         \\
                         &        &    43             &   \citet{lopezcorredoira2007}         \\
                         &        &   (42.44$\pm$2.14)  &   \citet{cabrera2008}         \\
                         &        &   $\left( 15^{+12.7}_{-13.3}\right) $  &   \citet{vanholle2009}         \\
                         &        &    $\sim$ 30    &   \citet{gonzalez2011}         \\
                         &        &         25 -- 27         &   \citet{nataf2012}         \\
                         &        &         13         &   \citet{robin2012}         \\
                         &        &         20         &   \citet{wang2012}         \\
                         &        &        29 -- 32         &   \citet{cao2013}         \\
 \enddata
 \tablenotetext{*}{Bulge following a truncated power law.}
 \tablenotetext{a}{Axis ratio 1.0:0.6:0.4 and angle between the Sun-center line and the major axis of the bulge $\sim$ 20$^{\circ}$.}
 \tablenotetext{b}{Axis ratio 1.00:0.68$^{+0.19}_{-0.05}$:0.31$^{0.04}_{-0.06}$
 and angle between the Sun-center line and the major axis of the bulge (15$^{+12.7}_{-13.3}$) deg.}
 \end{deluxetable}




\end{document}